\documentclass[aps,prl,twocolumn,nofootinbib,longbibliography,10pt]{revtex4}
\usepackage{etoolbox}
\usepackage{dcolumn}
\usepackage{amsmath,amssymb,amsfonts,mathtools}
\usepackage{mathrsfs}
\usepackage{graphicx}
\usepackage[colorlinks=true,urlcolor=blue,citecolor=red,linkcolor=blue]{hyperref}
\usepackage{natbib}
\usepackage{wrapfig}
\usepackage{flushend}

\begin{document}
\title{Minimal length scale correction in the noise of gravitons}
\author{Soham Sen}
\email{sensohomhary@gmail.com}
\affiliation{Department of Astrophysics and High Energy Physics, S. N. Bose National Centre for Basic Sciences, JD Block, Sector-III, Salt Lake City, Kolkata-700 106, India}
\author{Sunandan Gangopadhyay}
\email{sunandan.gangopadhyay@gmail.com}
\affiliation{Department of Astrophysics and High Energy Physics, S. N. Bose National Centre for Basic Sciences, JD Block, Sector-III, Salt Lake City, Kolkata-700 106, India}
\begin{abstract}
\noindent In this paper we have considered a quantized and linearly polarized gravitational wave interacting with a gravitational wave detector (interferometer detector) in the generalized uncertainty principle (GUP) framework. Following the analysis in \href{https://link.aps.org/doi/10.1103/PhysRevLett.127.081602}{Phys. Rev. Lett. 127 (2021) 081602}, we consider a quantized gravitational wave interacting with a gravitational wave detector (LIGO/VIRGO etc.) using a path integral approach. Although the incoming gravitational wave was quantized, no Planck-scale quantization effects were considered for the detector in earlier literatures. In our work, we consider a modified Heisenberg uncertainty relation with a quadratic order correction in the momentum variable between the two phase space coordinates of  the detector.
Using a path integral approach, we have obtained a stochastic equation involving the separation between two point-like objects. It is observed that random fluctuations (noises) and the correction terms due to the generalized uncertainty relation plays a crucial role in dictating such trajectories. Finally, we observe that the solution to the stochastic equation leads to time dependent standard deviation due to the GUP insertion, and for a primordial gravitational wave (where the initial state is a squeezed state) both the noise effect and the GUP effects exponentially enhance which may be possible to detect in future generation of gravitational wave detectors. We have also given a plot of the dimensionless standard deviation with time depicting that the GUP effect will carry a distinct signature which may be detectable in the future space based gravitational wave observatories. 

\end{abstract}
\maketitle
\noindent\textit{\textbf{Introduction: }}
It is a well known fact that a single particle freely falling under the effect of gravity, follows the geodesic equation, and in the case of a pair of point particles, they follow the geodesic deviation equation. These trajectories of freely falling objects under the effect of gravity is deterministic in nature and same follows for the geodesic deviation equation. These facts have been verified experimentally at a classical level, however their status at the quantum level has been the subject of intense research theoretically. Although the general theory of relativity gives a perfect description of gravity at a macroscopic level, the quantum nature of gravity is still unknown to the physics community. It is expected that gravity must have a quantum nature and therefore the search for a quantum theory of gravity is pursued. Recently there has been a very important work \cite{QGravNoise, QGravLett, QGravD} which demonstrated the effect on falling bodies due to the quantization of the gravitational field. It is observed that the dynamics of separation of a pair of falling bodies have a probabilistic nature which is different from the deterministic nature observed in the case of classical Einstein's gravity. It is also observed that the separation of two particles follows a Langevin-like stochastic equation with a random fluctuation term which is claimed as the quantum generalization of the classical geodesic deviation equation. The investigation reveals that linearized quantum theory of gravity can indeed have a startling effect on the motion of falling bodies. In this regard, we would like to stress that low energy quantum gravity corrections have been predicted by all the existing candidates for a quantum theory of gravity, such as string theory \cite{String1, String2}, loop quantum gravity \cite{Loop1, Loop2}, and noncommutative geometry \cite{Snyder,Snyder2,AC,NCG1}.  These theories indicate the presence of an observer-independent fundamental length scale (the Planck length $\approx10^{-33}$ m) in nature. However, for some string theories with dynamical string tensions, the existence of a fundamental length scale may be avoided. For example, in \cite{Guendelman}, a minimum time in the Euclidean space has been avoided, which would in principle be applicable to any spatial minimum length as well. By modifying the Heisenberg uncertainty principle (HUP), one can incorporate this minimal length. This modified  HUP is known as the generalized uncertainty principle (GUP). The relation between gravity and this minimal length scale was first shown in \cite{BRON1, BRON2} and later in \cite{MEAD}. In the gedanken experiments, a very strong signature of the existence of this minimal length scale was obtained. There have also been several investigations in different areas of theoretical physics to exploit this GUP framework, such as black hole physics \cite{MAGGIORE, SCARDIGLI, ADLERSAN, ADLERCHENSAN, RABIN, SG1, SCARDIGLI2, SG2, Ong, EPJC, BMajumder}, harmonic oscillators \cite{DAS1, DAS2}, optomechanical systems \cite{IVASP, IVASP2, KSP, OTM0}, and gravitational wave bar detectors \cite{SG3, SG4, OTM}. Recently there have been a few efforts to construct a laboratory-based test to investigate the effects of GUP in optomechanical systems\cite{Petruzzeillo, DasModak}. 

\noindent The general structure of the modified uncertainty relation is given by the following relation \cite{Farag, Farag2, SGSB} 
\begin{equation}\label{I.00}
\Delta \tilde{\xi}_i\Delta \tilde{\pi}_i\geq \frac{\hbar}{2}\left[1+\gamma \left(\Delta \tilde{\pi}^2+\langle \tilde{\pi}\rangle^2\right)+2\gamma \left(\Delta \tilde{\pi}_i^2+\langle \tilde{\pi}_i\rangle^2\right)\right]
\end{equation}
where the index $i$ runs from $1$ to $3$, and $\tilde{\xi}_i$ and $\tilde{\pi}_i$ are the phase space position and the conjugate momenta. The GUP parameter $\gamma$ in terms of the dimensionless parameter $\gamma_0$ can be recast as $\gamma=\frac{\gamma_0}{m_p^2c^2}$, where $m_p$ denotes the Planck mass and $c$ denotes the speed of light.  The GUP modified variables $\{\tilde{\xi}_i,\tilde{\pi}_i\}$ satisfies the following  commutation relation consistent with the uncertainty product given in eq.(\ref{I.00})
\begin{equation}\label{I.01}
[\tilde{\xi}_i,\tilde{\pi}_j]=i\hbar\left[\delta_{ij}+\gamma\left(\delta_{ij}\tilde{\pi}^2+2\tilde{\pi}_i\tilde{\pi}_j\right)\right]~.
\end{equation}
The above commutation relation reduces to the following form in one spatial dimension \cite{Farag,Farag2,SGSB,Kempf}
\begin{equation}\label{I.02}
[\tilde{\xi},\tilde{\pi}]=i\hbar[1+3\gamma \tilde{\pi}^2]~.
\end{equation} 
Following from the above commutation relation, the GUP modified variables $(\tilde{\xi},\tilde{\pi})$ in terms of the variables $(\xi,\pi)$  obeying the Heisenberg's uncertainty principle can be expressed as
\begin{equation}\label{I.1}
\tilde{\xi}=\xi~,~\tilde{\pi}=\pi(1+\gamma \pi^2)~.
\end{equation}
It is important to note that the coordinates  $\{\tilde{\xi},\tilde{\pi}\}$ obeying the modified commutation relation (eq.(\ref{I.02})) are being represented in terms of the coordinates $\{\xi,\pi\}$ which obey the usual commutation relation $[\xi,\pi]=i\hbar$.
Owning to the above discussion, it is natural to carry out the analysis in \cite{QGravNoise,QGravLett,QGravD} in the GUP framework. This would then incorporate effects from both the quantization of linearized gravity and modification of the Heisenberg's uncertainty relation.
\noindent The model considered in \cite{QGravNoise,QGravLett,QGravD} is as follows. The mirrors in the arm of the falling particles are considered as two freely falling particles. One of the particles has a heavier mass and the other particle has a comparatively smaller mass. The incoming gravitational wave is treated  quantum mechanically, and a perturbative approach is taken to include the quantum effects. Our analysis is significantly different from the previous analysis \cite{QGravNoise, QGravLett, QGravD} in the sense that we have now considered quantum gravity effects in the detector arm also by considering the existence of a minimal length scale. In order to truly incorporate the effects of the minimal length scale (having a direct connection to gravity), we have replaced the phase space coordinates of the detector part in the particle-graviton interaction Hamiltonian by GUP modified phase space coordinates. In order to analyze the system, we have considered a path integral approach and calculated the Feynman-Vernon influence functional \cite{FeynmanVernon} for the modified Hamiltonian. The incoming states have now been considered as minimum uncertainty states or coherent states which most closely resemble the classical gravitational wave. In our analysis, we observe as in \cite{QGravNoise,QGravLett,QGravD} that the separation between the two particles obeys a Langevin-like stochastic equation. However, we find terms containing both the effects of the modified uncertainty relation and the quantization of the gravitational waves. It is also observed that in the modified geodesic deviation equation, instead of second order time derivatives of the stochastic term, one can also observe first order time derivatives of the stochastic term solely due to the consideration of the modified uncertainty relation. We at first considered the graviton to exist in a coherent state and then we have extended our analysis for gravitons in squeezed states. In principle, one can find that the noise in the case of coherent states is very small but if one considers squeezed vacuum states, by tuning the squeezing parameter one may really detect such quantum gravitational effects in the next generation of gravitational wave detectors which could in principle give us hints about both the generalized uncertainty relation and the gravitons. Although having a squeezed vacuum state also have the limitation that there must be a source from which such gravitons in squeezed states are being generated. It is though assumed that primordial gravitational waves can be considered to exist in such squeezed states which can be a future candidate for the detection of such enhanced noise spectrum. At this point we would like to mention that inclusion of a minimal length scale with unmodified uncertainty principle and dispersion relation has also been reported in the literature \cite{Douglas}. However, the standard way of incorporating a minimal length scale is through the introduction of a GUP \cite{Kempf} or through a noncommutative algebra \cite{AC,Snyder,Snyder2}. In this work, we have followed this path.  
\vskip 0.2 cm
\noindent \textit{\textbf{Background: }}The main goal of this paper is to quantize the arm length of a gravitational wave detector (with arm length $\tilde{\xi}$) where the phase space coordinates of the detector follow the modified Heisenberg uncertainty priciple. The complete action of the system can be obtained by combining the Einstein-Hilbert action with the action of the detector. In terms of the GUP modified variables one can obtain the action of the system (considering only one polarization of the gravitational wave) as 
\begin{equation}\label{action}
S_{\mathcal{A}}=\int dt\left(\frac{m}{2}\left(\dot{q}^2-\omega^2q^2\right)+\frac{m_0}{2}\dot{\tilde{\xi}}^2-\mathcal{G}\dot{q}\dot{\tilde{\xi}}\tilde{\xi}\right)
\end{equation}
 with $q$ denoting the configuration space variable of the gravitational wave, $\hbar\omega$ being the energy of the gravitational wave mode, and $m_0$ denoting the smaller mass between the two masses representing the interferometer. Here, $\mathcal{G}=\frac{m_0}{2 l_p}$ denotes the graviton-detector coupling constant with $l_p=\sqrt{\frac{\hbar G}{c^3}}$ being the Planck length and $m=\frac{L^3c^5}{16\hbar G^2}$. 
The Hamiltonian from the action in terms of the phase space variables of the gravitational wave ($\{q,p\}$) and the GUP modified phase space variables of the gravitational wave detector ($\{\tilde{\xi},\tilde{\pi}\}$) is given by 
\begin{equation*}
 H=\left(\frac{p^2}{2m}+\frac{\tilde{\pi}^2}{2m_0}+\frac{\mathcal{G}p\tilde{\pi}\tilde{\xi}}{mm_0}\right)\left[1-\frac{\mathcal{G}^2\tilde{\xi}^2}{mm_0}\right]^{-1}+\frac{1}{2}m\omega^2q^2~.
\end{equation*}
The above Hamiltonian in terms of the unmodified detector variables ($\{\xi,\pi\}$) upto $\mathcal{O}(\gamma)$ can be recast as 
\begin{equation}\label{1.1}
H=\frac{\frac{p^2}{2m}+\frac{\pi^2}{2m_0}+\frac{\mathcal{G}p\pi\xi}{mm_0}+\gamma\left(\frac{\pi^4}{m_0}+\frac{\mathcal{G}p\pi^3\xi}{mm_0}\right)}{1-\frac{\mathcal{G}^2\xi^2}{mm_0}}+\frac{1}{2}m\omega^2q^2
\end{equation}
where we have used $\tilde{\pi}=\pi(1+\gamma \pi^2)$. Here, $\xi$ denotes the geodesic separation of the lighter mass $m_0$ from the heavier mass. One can now elevate this Hamiltonian from classical description to quantum description by introducing appropriate commutation relations between the coordinate variables and their conjugate momenta. In this analysis we consider the initial state of the gravitational wave as $|\psi_\omega\rangle$ and that of the final state as $|\mathcal{F}\rangle$. As we do not know the final state of the gravitational wave, we need to sum over all $\lvert\mathcal{F}\rangle$ states. As initially there was no coupling between the initial gravitational wave state and the detector state, we will consider them as a tensor product state. Now via spontaneous and stimulated emission procedures the detector masses will both emit and absorb gravitons. Here, our aim is to compute the probability and the form of the probability is given as follows
\begin{equation}\label{1.2}
P_{\psi_{\omega}}^{[\phi_i\rightarrow\phi_f]}=\sum\limits_{\lvert\mathcal{F}\rangle}\left|\langle \mathcal{F},\phi_f|\hat{U}(T+\Delta,-\Delta)|\psi_\omega,\phi_i\rangle\right|^2
\end{equation} 
where $|\phi_f\rangle$ and $|\phi_i\rangle$ are the final state and the initial states of the particle at times $t=T+\Delta$ and $t=-\Delta$ respectively. $\hat{U}$ is the unitary time evolution operator in eq.(\ref{1.2}) and associated with the quantum mechanical analogue to the Hamiltonian in eq.(\ref{1.1}). It is very important to note that the interaction is turned on in the interval $t=0$ to $t=T$. Inserting complete set of joint position eigenstates and summing over all the final gravitational wave states $|\mathcal{F}\rangle$ in eq.(\ref{1.2}), one can rewrite the transition probability as 
\begin{equation}\label{1.3}
\begin{split}
&P_{\psi_{\omega}}^{[\phi_i\rightarrow\phi_f]}=\int dq_idq_i'dq_fd\xi_id\xi_i'd\xi_fd\xi_f'\psi_\omega(q_i)\psi^*_\omega(q_i')\phi_i(\xi_i)\\
&\times\phi_i^*(\xi'_i)\phi_f^*(\xi_f)\phi_f(\xi_f')~\langle q_i',\xi_i'|\hat{U}^\dagger(T+\Delta,-\Delta)|q_f,\xi_f'\rangle\\
&\times\langle q_f,\xi_f|\hat{U}(T+\Delta,-\Delta)|q_i,\xi_i\rangle~.
\end{split}
\end{equation}
Rewriting each of the amplitudes present in the canonical path integral form and executing the path integral over $\pi$, one can now recast the amplitude in a much compact form given by 
\begin{equation*}
\begin{split}
&\langle q_f,\xi_f|\hat{U}(T+\Delta,-\Delta)|q_i,\xi_i\rangle=\int \tilde{\mathcal{D}}\xi \exp\Bigr[\frac{im_0}{2\hbar}\int_{-\Delta}^{T+\Delta}dt\\&\times [\dot{\xi}^2-2\gamma m_0^2\dot{\xi}^4]\Bigr]\int\mathcal{D}q\mathcal{D}pe^{\frac{i}{\hbar}\int_{-\Delta}^{T+\Delta}dt[p\dot{q}-H^\gamma_{\xi}[q,p]]}~.  
\end{split}
\end{equation*}
In the above expression, the form of the reduced Hamiltonian is given by 
\begin{equation*}
H^\gamma_{\xi}[q,p]=\frac{(p+\mathcal{G}\xi\dot{\xi}(1-3\gamma m_0^2\dot{\xi}^2))^2}{2m}+\frac{1}{2}m\omega^2q^2~.
\end{equation*}
We can finally obtain the form of the probability in eq.(\ref{1.3}) as
\begin{equation}\label{1.4}
\begin{split}
&P_{\psi_{\omega}}^{[\phi_i\rightarrow\phi_f]}=\int d\xi_id\xi_i'd\xi_fd\xi_f'\phi_i(\xi_i)\phi_i^*(\xi'_i)\phi_f^*(\xi_f)\phi_f(\xi_f')\\&\int\tilde{\mathcal{D}}\xi\tilde{\mathcal{D}}\xi'e^{\frac{i}{\hbar}\int_{-\Delta}^{T+\Delta}dt\frac{m_0}{2}\left[(\dot{\xi}^2-\dot{\xi}'^2)-2\gamma m_0^2(\dot{\xi}^4-\dot{\xi}'^4)\right]}F^{\gamma}_{\psi_\omega}[\xi,\xi']
\end{split}
\end{equation}
where we have defined the following two quantities, 
\begin{align*}
&\langle q_f| \hat{U}^\gamma_{\xi}(T+\Delta,-\Delta)|q_i\rangle\equiv\int\mathcal{D}q\mathcal{D}pe^{\frac{i}{\hbar}\int_{-\Delta}^{T+\Delta}dt[p\dot{q}-H^\gamma_{\xi}[q,p]]},\\
&F^{\gamma}_{\psi_\omega}[\xi,\xi']\equiv\langle \psi_\omega|\hat{U}^{\gamma\dagger}_{\xi'}[T+\Delta,-\Delta]\hat{U}^\gamma_{\xi}[T+\Delta,-\Delta]|\psi_\omega\rangle~.
\end{align*}
Here, $F^{\gamma}_{\psi_\omega}[\xi,\xi']$ gives the Feynman-Vernon influence functional. It is very important to observe that influence functional is the only term in the probability containing the harmonic oscillator state ($|\psi_\omega\rangle$) and providing a coupling between the harmonic oscillator and the detector variable. With the generic structure of the probability in hand, we shall now proceed to explicitly compute the form of the influence functional.  It is important to observe that we are currently running our analysis for a single mode of the gravitational wave and we shall extend it for multiple modes also.
\vskip 0.2 cm
\noindent \textit{\textbf{The influence functional: }}From the form of the Hamiltonian $\hat{H}^{\gamma}_\xi(q,p)$, it can be seen that the instantaneous eigenstates are general harmonic oscillator eigenstates generated by the shift of the momentum in the momentum space via a parameter $\mathcal{G}\xi\dot{\xi}(1-3\gamma m_0^2\dot{\xi}^2)$. In order the convert the the integrals from $-\Delta $ to $T+\Delta$ to integrals from $0$ to $T$, we need to redefine the Heisenberg eigenstates from $|\psi_\omega\rangle$ to  $e^{-\frac{i}{\hbar}\hat{H}_0\Delta}|\psi_\omega\rangle$ and we also need to introduce the modified wave functions given by the following two relations 
\begin{align*}
\tilde{\phi}_i(\tilde{\xi}_i)&=\int d\xi_i\phi_i(\xi_i)\int\tilde{\mathcal{D}}\xi e^{\frac{i}{\hbar}\int_{-\Delta}^0dt\frac{m_0}{2}(\dot{\xi}^2-2\gamma m_0^2\dot{\xi}^4)}~,\\
\tilde{\phi}_f(\tilde{\xi}_f)&=\int d\xi_f\phi_f(\xi_f)\int\tilde{\mathcal{D}}\xi e^{-\frac{i}{\hbar}\int_{T}^{T+\Delta}dt\frac{m_0}{2}(\dot{\xi}^2-2\gamma m_0^2\dot{\xi}^4)}~.
\end{align*}
 In the above two relations, we have defined $\xi(-\Delta)=\xi_i$, $\xi(T+\Delta)=\xi_f$, $\xi(0)=\tilde{\xi}_i$, and $\xi(T)=\tilde{\xi}_f$. The modified probability formula takes the form as follows
\begin{equation}\label{1.5}
\begin{split}
&P_{\psi_{\omega}}^{[\phi_i\rightarrow\phi_f]}\equiv\int d\xi_id\xi_i'd\xi_fd\xi_f'\phi_i(\xi_i)\phi_i^*(\xi'_i)\phi_f^*(\xi_f)\phi_f(\xi_f')\\&\int[\tilde{\mathcal{D}}\xi]_{\xi_i,0}^{\xi_f,T}[\tilde{\mathcal{D}}\xi']_{\xi'_i,0}^{\xi'_f,T}e^{\frac{i}{\hbar}\int_{0}^{T}dt\left[L_0^\gamma(\xi)-L_0^\gamma(\xi')\right]}F^{\gamma}_{\psi_\omega}[\xi,\xi']
\end{split}
\end{equation}
where $L_0^\gamma(\xi)=\frac{m_0}{2}(\dot{\xi}^2-2\gamma m_0^2\dot{\xi}^4)$ and we have redefined the detector variable from $\tilde{\xi}$ to $\xi$ (with $\xi(0)=\xi_i$, $\xi(T)=\xi_f$ and $\xi'(0)=\xi'_i$, $\xi'(T)=\xi'_f$). Due to the change in the integration limit, the Feynman-Vernon influence functional also gets modified. The modified Feynman-Vernon influence functional has the form 
\begin{equation}
\begin{split}
&F^\gamma_{\psi_\omega}[\xi,\xi']=\langle \psi_\omega|e^{\frac{i}{\hbar}\hat{q}\mathcal{G}\xi_i'\dot{\xi}'_i[1-3\gamma m_0^2\dot{\xi}'^2_i]}\hat{U}^{I\dagger}_{\gamma\xi'}(T,0)\\&\times e^{-\frac{i}{\hbar}\hat{q}_I(T)\mathcal{G}\xi_f'\dot{\xi}'_f[1-3\gamma m_0^2\dot{\xi}'^2_f]} e^{\frac{i}{\hbar}\hat{q}_I(T)\mathcal{G}\xi_f\dot{\xi}_f[1-3\gamma m_0^2\dot{\xi}^2_f]}\\&\times\hat{U}^{I}_{\gamma\xi}(T,0)e^{-\frac{i}{\hbar}\hat{q}\mathcal{G}\xi_i\dot{\xi}_i[1-3\gamma m_0^2\dot{\xi}^2_i]}|\psi_\omega\rangle
\end{split}
\end{equation}
where $\hat{U}^{I}_{\gamma\xi}(T,0)$ gives the unitary time evolution operator in the interaction picture and $\hat{q}_I(T)=e^{\frac{i}{\hbar}\hat{H}_0T}\hat{q}e^{\frac{i}{\hbar}\hat{H}_0T}$. In order to proceed further, we need to decompose the unitary time evolution operators in terms of the time ordered exponential functions in the interaction picture and finally write the form of the unitary time evolution operator in the interaction picture as follows
\begin{equation}\label{1.6}
\begin{split}
&\hat{U}^{I}_{\gamma\xi}(T,0)\equiv e^{-\frac{i\mathcal{G}}{\hbar}\hat{q}_I(T)\xi_f\dot{\xi}_f[1-3\gamma m_0^2\dot{\xi}_f^2]} e^{\frac{i\mathcal{G}}{2\hbar}\int_0^Tdt~\hat{q}_I(t)Z(t)}\\&e^{\frac{i\mathcal{G}}{\hbar}\hat{q}_I(0)\xi_i\dot{\xi}_i(1-3\gamma m_0^2\dot{\xi}_i^2)}e^{-\frac{\mathcal{G}^2}{8\hbar^2}\int_0^Tdt\int_0^tdt'Z(t)Z(t')[\hat{q}_I(t),\hat{q}_I(t')]}
\end{split}
\end{equation}
where $X(t)=\frac{d^2}{dt^2}(\xi^2)$ and $Y(t)=\frac{d}{dt}(\dot{\xi}^2\frac{d}{dt}(\xi^2))$, and $Z(t)=X(t)-3\gamma m_0^2 Y(t)$. By the repeated use of the Baker-Campbell-Hausdorff formula and substituting the form of the unitary time evolution operator from eq.(\ref{1.6}) into the Feynman-Vernon influence functional, we obtain the modified form of the influence functional as follows
\begin{equation}\label{1.7}
\begin{split}
F^\gamma_{\psi_\omega}[\xi,\xi']=\langle \psi_\omega |e^{-V^*\hat{a}^\dagger}e^{V\hat{a}}|\psi_\omega\rangle e^{i\Phi^\gamma_{0_\omega}[\xi,\xi']}
\end{split}
\end{equation}
where
\begin{equation}\label{1.7a}
V=\frac{i\mathcal{G}}{\sqrt{8\hbar m\omega}}\int_0^Tdt e^{-i\omega t}(Z(t)-Z'(t))
\end{equation}
 and the phase factor in the exponential term of eq.(\ref{1.7}) is given by the following relation 
\begin{equation}\label{1.7b} 
\begin{split}
 i\Phi^\gamma_{0_\omega}[\xi,\xi']\equiv&-\frac{\mathcal{G}^2}{8\hbar m\omega}\int_0^Tdt\int_0^t dt'(Z(t)-Z'(t))\\&\times(Z(t')e^{-i\omega(t-t')}-Z'(t')e^{i\omega(t-t')})~.
 \end{split}
\end{equation}
With the form of the Feynman-Vernon influence functional in hand, we are now in a position to consider different cases for the state $|\psi_\omega\rangle$ of the gravitational wave. 
We now consider a gravitational wave-mode in a coherent state $|\psi_\omega\rangle=|\alpha_\omega\rangle$  with eigenvalue $\alpha_\omega=\sqrt{\frac{m\omega}{2\hbar}}\zeta_\omega e^{-i\phi_\omega}$ where the form of the classical gravitational wave mode is given by $q_{cl}(t)=\zeta_\omega \cos(\omega t+\phi_\omega)$. One can now easily compute the form of the influence functional in the single-mode analysis and proceed to compute the influence functional for a continuum of such modes (or a gravitational field). The influence functional of the field can be considered as a product of the influence functional for the individual modes ($F^\gamma_{\Psi}[\xi,\xi']=\prod\limits_{\vec{k}}F^\gamma_{\psi_{\omega(\vec{k})}}[\xi,\xi']$) where the gravitational field is given as $|\Psi\rangle=\mathop{\otimes}\limits_{\vec{k}}|\psi_{\omega(\vec{k})}\rangle$. One can finally compute the transition probability for a gravitation field with its field modes in coherent states as follows
\begin{equation}\label{1.8}
\begin{split}
&P_{\Psi}\equiv \int d\xi_id\xi'_id\xi_fd\xi'_f\phi_i(\xi_i)\phi_i^*(\xi_i')\phi_f^*(\xi_f)\phi_f(\xi_f')\int\tilde{\mathcal{D}}\xi\tilde{\mathcal{D}}\xi'\\
&\times\int\tilde{\mathcal{D}}\mathcal{N}_0~\exp\left[-\frac{1}{2}\int_0^Tdt\int_0^Tdt'\mathcal{A}_0^{-1}(t,t')\mathcal{N}_0(t)\mathcal{N}_0(t')\right]\\
&\times \exp\biggr[\frac{im_0}{2\hbar}\int_0^T dt\biggr[(\dot{\xi}^2-\dot{\xi'}^2)-2\gamma m_0^2(\dot{\xi}^4-\dot{\xi'}^4)\\&+\left[\frac{(\bar{h}(t)+\mathcal{N}_0(t))}{2}-\frac{m_0G}{4}[\dot{Z}(t)+\dot{Z}'(t)]\right][Z(t)-Z'(t)]\biggr]
\end{split} 
\end{equation}
where $\bar{h}(t)=\frac{1}{l_p}\sum\limits_{\omega}\zeta_\omega\cos(\omega t+\phi_\omega)$. To obtain the final form of the probability in eq.(\ref{1.8}), we have made use of the Feynman-Vernon trick. The function $\mathcal{N}_0(t)$ has the interpretation of a noise term, that is a stochastic random function with a Gaussian probability density. Indeed, one can define the average of $\mathcal{N}_0(t)$ which vanishes \cite{QGravD}
\begin{align}
\langle\mathcal{N}_0(t)\rangle&\equiv\int\mathcal{D}\mathcal{N}_0\exp\Bigr[-\frac{1}{2}\int_0^T\int_0^Tdt' dt''\mathcal{A}_0^{-1}(t',t'')\nonumber\\&\times\mathcal{N}_0(t')\mathcal{N}_0(t'')\Bigr]\mathcal{N}_0(t)=0\label{1.8a}~.
\end{align}
The function $\mathcal{A}_0(t,t')$ is the autocorrelation function of $\mathcal{N}_0(t)$ as \cite{QGravD}
\begin{align}
&\langle \mathcal{N}_0(t)\mathcal{N}_0(t')\rangle\equiv \int\mathcal{D}\mathcal{N}_0\exp\Bigr[-\frac{1}{2}\int_0^T\int_0^Tdt'' dt'''\nonumber\\&\times\mathcal{A}_0^{-1}(t'',t''')\mathcal{N}_0(t'')\mathcal{N}_0(t''')\Bigr]\mathcal{N}_0(t)\mathcal{N}_0(t')=\mathcal{A}_0(t,t')~.\label{1.8b}
\end{align}
This autocorrelation function will be of immense importance as we shall see below. In case of the coherent state analysis, the auto correlation function has the form \cite{QGravD}
\begin{equation}\label{1.8c}
\mathcal{A}_0(t,t')=\frac{4\hbar G}{\pi}\int_0^\infty d\omega\omega \cos(\omega(t-t'))
\end{equation} 
which is in general divergent in nature. But for a gravitational wave detector, it is sensitive to a certain range of the gravitational wave frequency and therefore one can regularize the integral by applying a maximum cut-off frequency $\omega_{max}$. Due to the usage of a dipole-like approximation in our current analysis, we can use the maximum value of the frequency as $\frac{2\pi c}{\xi_0}$ with $\xi_0$ being the resting arm length of the detector.
\vskip 0.2 cm
\noindent\textbf{\textit{Dynamics of the arm length: }}With the form of the probability in hand, we can now compute the quantum-dynamics of the detector arm length $\xi$. In order to obtain the effective stochastic equation, we consider the saddle point approximation. The saddle point gives the maximum contribution
to the path integral in eq.(\ref{1.8}). For the gravitational wave in the coherent state, following the procedure in \cite{QGravD}, one can obtain the following differential equation for the detector variable $\xi(t)$
\begin{equation}\label{1.9}
\begin{split}
&\ddot{\xi}-\frac{1}{2}\biggr[\left(\ddot{\bar{h}}(t)+\ddot{\mathcal{N}}_0(t)-\frac{m_0G}{c^5}\frac{d^5}{dt^5}(\xi^2)\right)(1+3\gamma m_0^2\dot{\xi}^2)+\\&\frac{3\gamma m_0^3G }{c^5}\frac{d^4}{dt^4}\left(\frac{d}{dt}(\xi^2)\dot{\xi}^2\right)\biggr]\biggr]\xi+3\gamma m_0^2(\dot{\xi}^3+3\xi\dot{\xi}\ddot{\xi})\biggr[\dot{\bar{h}}(t)+\\&\dot{\mathcal{N}}_0(t)-\frac{m_0G}{c^5}\frac{d^4}{dt^4}(\xi^2)\biggr]=0~.
\end{split}
\end{equation}
Eq.(\ref{1.9}) is one of the main results in our paper. It is important to observe that the geodesic deviation equation is replaced now by a quantum stochastic-equation. The most important observation is that now the quantum geodesic equation is governed by the terms coupling the effects of the GUP parameter and the noise term. Unlike \cite{QGravNoise,QGravLett,QGravD}, there are also terms involving the first order derivative of the noise term with respect to time. The $\ddot{\bar{h}}(t)\xi$ term in eq.(\ref{1.9}) is a tidal acceleration term due to the passing of a classical gravitational wave and the fifth order time derivative term is the dissipative gravitational radiation reaction term \cite{KSThorne,WLBurke,Chandrasekhar,Sasaki}. This term $\ddot{\bar{h}}(t)\xi$ will be important as will be clear in the subsequent discussion. The other higher derivative terms are corrections to the gravitational radiation reaction due to the modification in the uncertainty relation of the detector variables.

\noindent For a gravitational wave in the coherent state, the noise spectrum is of the order of the Planck length making it a near impossible task to detect the signatures of the gravitons, let alone the generalized uncertainty relation. Therefore, we shall consider the gravitational wave in a squeezed state $\hat{S}_{z_{\omega}}|0_\omega\rangle$, where $\hat{S}_{z_{\omega}}=e^{\frac{1}{2}(z_\omega^{*}\hat{a}^2-z_\omega \hat{a}^{\dagger2})}$ gives the squeezing operator with $z_\omega=r_\omega e^{i\phi_\omega}$ being the squeezing parameter. The quantum geodesic equation for a gravitational wave in a squeezed state gives the following equation
\begin{equation}\label{1.10}
\begin{split}
&\ddot{\xi}-\frac{1}{2}\biggr[\left(\ddot{\mathcal{N}}^{n.s.}(t)+\sqrt{\cosh 2r}\ddot{\mathcal{N}}_0(t)-\frac{m_0G}{c^5}\frac{d^5}{dt^5}(\xi^2)\right)(1+\\&3\gamma m_0^2\dot{\xi}^2)+\frac{3\gamma m_0^3G }{c^5}\frac{d^4}{dt^4}\left(\frac{d}{dt}(\xi^2)\dot{\xi}^2\right)\biggr]\biggr]\xi+3\gamma m_0^2\biggr[\dot{\mathcal{N}}^{n.s.}(t)\\&+\sqrt{\cosh 2r}\dot{\mathcal{N}}_0(t)-\frac{m_0G}{c^5}\frac{d^4}{dt^4}(\xi^2)\biggr](\dot{\xi}^3+3\xi\dot{\xi}\ddot{\xi})=0
\end{split}
\end{equation}
where $r$ denotes the real part of the squeezing parameter for the quantum field and $\mathcal{N}^{n.s.}(t)$ denotes the non-stationary noise generated due to the time  modulation of the noise in squeezed states. It is very important to observe in eq.(\ref{1.10}) that the associated noise term now can be exponentially increased along with the terms generated due to the generalized uncertainty principle for values of $r$ greater than unity. This induces an increased chance in the detection of such minuscule corrections present in such gravitational wave detection scenarios. To proceed further, it is important to note that we can obtain a solution of the given Langevin-like equations by means of perturbative calculations. We can get rid of the dissipative gravitational radiation reaction terms in eq.(s)(\ref{1.9},\ref{1.10}) \cite{QGravD}. In order to do so one needs to consider that $\xi$ is measured in a coarse grained manner which in turns result in the higher derivatives to be negligible. We shall try to obtain an approximate solution of the time dependent geodesic separation for the gravitational wave to be initially in a coherent state.  
In order to find a solution to eq.(s)(\ref{1.9},\ref{1.10}), we use an iterative approach. For the base equation $\ddot{\xi}(t)=0$ without the higher order terms, we can obtain a zeroth order solution of the form $\xi^{(0)}(t)=\xi_0+\lambda t$, where the constant $\lambda$ has the dimension of velocity and can have a maximum value $\lambda=c$. 
The maximum interaction time between the graviton being absorbed and released by the detector is $t_{\text{max}}\sim\frac{\xi_0}{c}$ and therefore the linear time dependent term in $\xi(t)$ can go up to $\lambda t_{\text{max}}$. Following the same iterative procedure, we can obtain a most general solution of eq.(\ref{1.9}) up to $\mathcal{O}(\gamma,\mathcal{N}_0,\bar{h},\gamma\mathcal{N}_0,\gamma\bar{h})$ as follows
\begin{align}
\xi(t)\cong&(\xi_0+\lambda t)\left[1+\frac{1}{2}\left[1+3\gamma m_0^2\lambda^2\right])(\bar{h}(t)+\mathcal{N}_0(t))\right]\nonumber\\&-\lambda(1+6\gamma m_0^2\lambda^2)\int_0^t dt'(\bar{h}(t')+\mathcal{N}_0(t'))~.\label{1.11}
\end{align}
 Now, the higher limit of the integral in eq.(\ref{1.11}) has a cut-off at $t=t_{\text{max}}$. 
Our aim now is to calculate the standard deviation $\sigma=\sqrt{\left<(\xi(t)-\left<\xi(t)\right>)^2\right>}$. We can separate the standard deviation term into two parts as $\sigma(t)\cong\sigma_0(t)+\sigma_{\gamma}(t)$. 
The $\sigma_0(t)$ part has been calculated in \cite{QGravLett,QGravD} and reads
\begin{equation}\label{1.11a}
\sigma_0(t)\sim\sqrt{2\pi}l_p\sim 10^{-35}\text{ m}~.
\end{equation}
 In case of the GUP contribution of the standard deviation, we need to consider the time dependent parts and ignore the linear time-dependent contribution. We then obtain the form of the dimensionless parameter$\frac{\sigma_\gamma(t)}{\sqrt{2\pi}l_p}$ to be (with $\lambda$ set to its maximum value)
\begin{equation}\label{1.12}
\begin{split}
&\frac{\sigma_\gamma(t)}{\sqrt{2\pi}l_p}\cong3\gamma m_0^2 c^2\biggr[1-\frac{2}{\pi(1+\frac{c t}{\xi_0})}\frac{\xi_0 \sin^2\left[\frac{\pi c t}{\xi_0}\right]}{\pi c t}\\&+\frac{4}{\pi^2 \left(1+\frac{c t}{\xi_0}\right)^2}\left(\gamma_\varepsilon-\text{Ci}\left[\frac{2\pi c t}{\xi_0}\right]+\ln\left[\frac{2\pi c t}{\xi_0}\right]\right)\biggr]
\end{split}
\end{equation}
where $\gamma_\varepsilon$ gives the Euler constant and $\text{Ci}$ denotes the cosine integral function \cite{Gradshteyn}.
For a gravitational wave in the squeezed coherent state, we obtain the form of the standard deviation to be (considering the static part only)
\begin{equation}\label{1.12a}
\sigma_{\text{Squeezed}}(t)=\sqrt{\cosh 2r}\sigma(t)
\end{equation}
 where $\sigma(t)=\sigma_0(t)+\sigma_\gamma(t)$. 
\vskip 0.2 cm
\noindent\textit{\textbf{Phenomenological aspects of the model: }} 
It is important to note that the gravitational wave observatories LIGO (/VIRGO) has an $L$ shaped structure with the arm length at rest to be $\xi_0=4 \text{ km}$ ($\xi_0=3\text{ km}$ for VIRGO). For the mirror suspended at the both ends of the Fabry-Perot cavity, the mirror coating is made up of fused Silica (mass of a single $\text{SiO}_2$ molecule is $m_{\text{SiO}_2}\sim10^{-25}\text{ kg}$) which serves as the low-index layer and Tantalum pentoxide (mass of a single $\text{Ta}_2\text{O}_5$ molecule is $m_{\text{Ta}_2\text{O}_5}\sim 10^{-24}\text{ kg}$) which serves as the high-index layer \cite{Accadia}. We can indeed obtain a bound on the GUP parameter using these parameters from the existing gravitational wave observatories. Note that, $\gamma m_0^2c^2\sim\gamma_0 \times 10^{-33}$ (for $m_0\sim10^{-24}\text{ kg}$) and from the requirement $\zeta\gamma m_0^2 c^2<1$ (where the dimensionless constant $\zeta$ is a number of order 10), we can impose a bound on the dimensionless quadratic GUP parameter to be $\gamma_0<10^{31}$ which is weaker than the bound obtained earlier for a resonant bar detector interacting with a gravitational wave in \cite{SukantaDaSunandan} but tighter than the bound obtained in \cite{GWaveDas} using gravitational wave observation data.
We shall now try to give a basic estimate on the detectability of the GUP effect from the standard deviation $\sigma(t)$. For the gravitational wave to be in coherent state, it is important to understand that $\sigma\cong \sqrt{2\pi}l_p\sim10^{-35} \text{m}$ and current detectability lies around $10^{-18}$ m. In case of the initial graviton state being in a squeezed state, we can indeed observe that $\sigma^{\text{Squeezed}}(t)=\sqrt{\cosh 2r}\sigma(t)$ which indicates that for a sufficiently high squeezing parameter the standard deviation due to the induced noise and the GUP effect may be detectable.  
\begin{figure}
\begin{center}
\includegraphics[scale=0.2915]{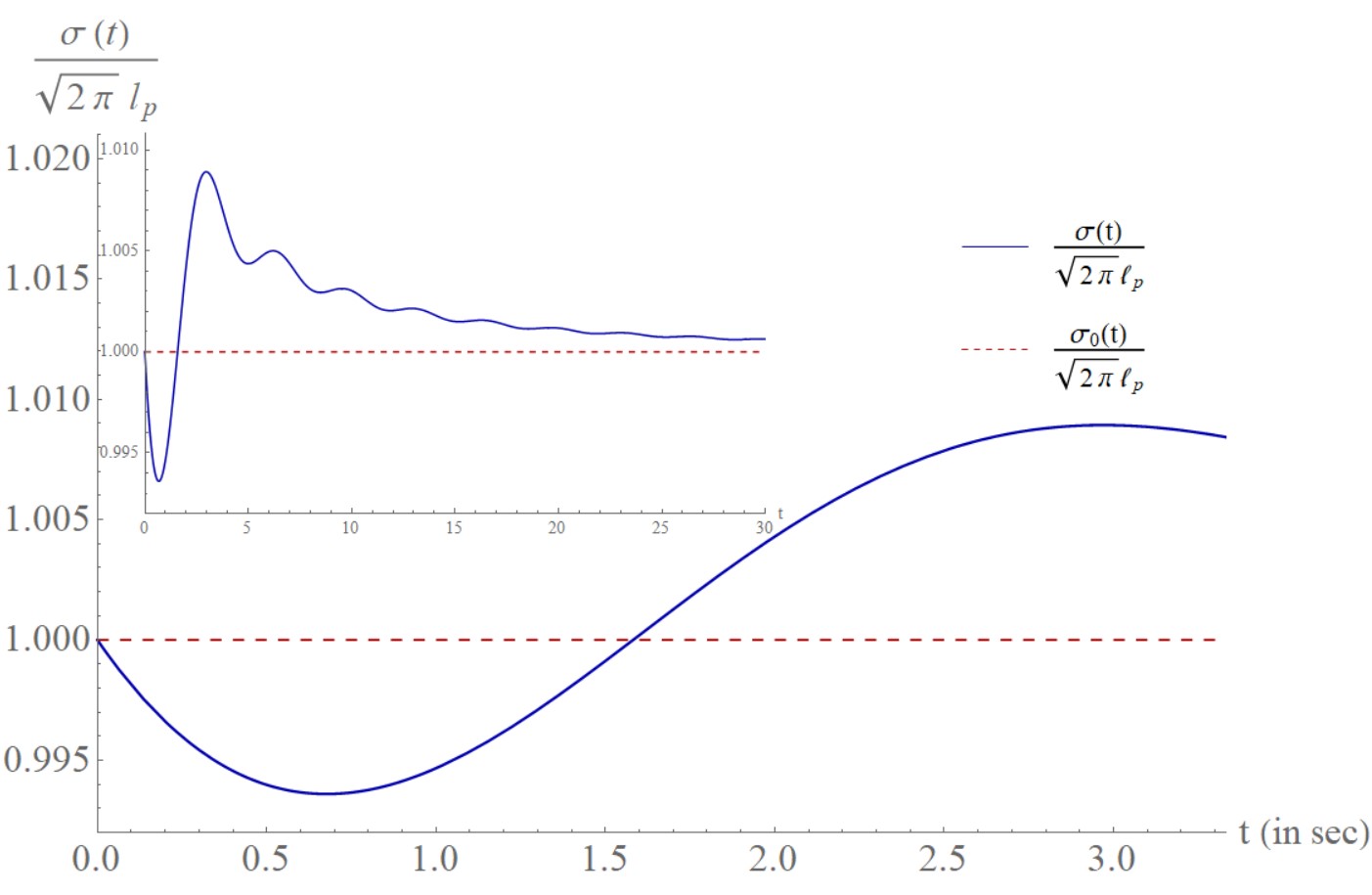}
\caption{$\frac{\sigma(t)}{\sqrt{2\pi}l_p}$ vs $t$ plot for gravitational wave with maximum frequency $\omega_{\text{max}}\sim 1 \text{ Hz}$\label{F1}. The inset image depicts the long term behaviour of the time dependent standar deviation and it signifies that the standard deviation asymptotes towards the value of the standard deviation obtained in \cite{QGravD,QGravLett} (dotted line) with increase in time $t$. }
\end{center}
\end{figure}
In general such states can only generate in post-inflationary scenarios \cite{PI1,PI2,PI3} leading to a very rare chance of detection of such  primordial gravitational waves.  For a ``grand unified theory" inflation, the frequency is at $\omega\sim 0.1 Hz$ which is beyond the frequency range of both LIGO and VIRGO. This frequency range can although be detectable by the future space based gravitational wave observatories DECIGO or LISA\footnote{DECIGO: Decihertz Interferometer Gravitational Wave Observatory and LISA: Laser Interferometer Space Antenna.}. For such a primordial gravitational wave $e^{r}\sim10^{18}$ \cite{Soda} which results in the enhancing parameter to be of the order of $\sqrt{\cosh 2r}\cong \frac{1}{\sqrt{2}}e^{r}\sim 10^{18}$. We will primarily consider the LISA interferometer with $\xi_0\sim10^6 \text{ km}$ and maximum frequency $\omega_{\text{max}}\sim 1\text{ Hz}$ \cite{QGravD}. LISA has a projected sensitivity at $10^{-18}$ m \cite{QGravD}. With the inherent squeezing in the primordial gravitational wave and a maximum interaction time of $t_{\text{max}}\sim 3.33 \text{ sec}$ one can estimate that the standard deviation is around (while the detector has just stopped interacting with the gravitational wave) $\sigma_0^{\text{Squeezed}}(t_\text{max})\sim \sqrt{\cosh 2r}\times 10^{-35} \text{ m}\sim 10^{-17}\text{ m}$ which is in the detectable range of the LISA observatory. A very important result can be observed in terms of the GUP part of the standard deviation. 
The contribution to the standard deviation due to the GUP effect can be observed (from eq.(\ref{1.12})) to have a value $\sigma_\gamma^{\text{Squeezed}}(t_{\text{max}})\sim \sqrt{\cosh 2r}~ 10^{-37}\text{ m}\sim 10^{-19}\text{ m}$. 
Hence, we find that for primordial gravitational waves, the standard deviation carrying the signature of the graviton has a value $\sigma_0^{\text{Squeezed}}\sim 10^{-17}\text{ m}$ and the GUP effect lies in the range $\sigma^{\text{Squeezed}}_{\gamma}\sim (10^{-19}-10^{-20})\text{ m}$ which is just one order of magnitude beyond the projected sensitivity of the LISA observatory. The ratio of $\sigma_\gamma^{\text{Squeezed}}(t_\text{max})$ with $\sigma_0^{\text{Squeezed}}(t_\text{max})$ is $\sigma_\gamma^{\text{Squeezed}}/\sigma_0^{\text{Squeezed}}\sim10^{-2}$. The analysis has been done for a primordial gravitational wave with squeezing parameter $r\sim 42$. If it is possible to detect a primordial gravitational wave with a squeezing parameter $r\sim 44$, we then find that the GUP contribution to have a value $\sigma^{\text{Squeezed}}_{\gamma}(t)\sim10^{-18}\text{ m}$ which indeed will be in the projected sensitivity range of the LISA observatory. It will then be possible to detect the stochastic noise effect due to the existence of gravitons as well as the existence of GUP. Plotting the standard deviation due to the GUP contribution divided by $\sqrt{2\pi} l_p$ with respect to time, we can observe a very unique behaviour based on the time of sampling of the $\xi(t)$ data in Fig.(\ref{F1}). We observe from Fig.(\ref{F1}) that the value of the dimensionless number $\frac{\sigma(t)}{\sqrt{2\pi}l_p}$ decreases and then increases with time.  Here, we have compared the result of the standard deviation obtained in our analysis with the result obtained in \cite{QGravD,QGravLett}. Here, $\sigma(t)=\sigma_0(t)+\sigma'_{\gamma}(t)$ with the definition $\sigma'_{\gamma}(t)\equiv\sigma_{\gamma}(t)-\sigma_{\gamma}(0)$.
Fig.(\ref{F1}) (and the small inset Figure depicting long time behaviour) suggests that the sudden increase in the value of the standard deviation decreases gradually with time and returns back to the  value of the standard deviation $\sigma_0(t)=\sqrt{2\pi }l_p$, given in \cite{QGravD,QGravLett}, in the long time limit. If the standard deviation can be calculated at a fixed time interval during when the interaction happens and such a dip in the standard deviation value is observed, we can claim the existence of the generalized uncertainty principle. Hence, for an advanced gravitational wave observatory, it may be possible to detect both the existence of gravitons as well as the existence of a fundamental minimal length scale correction in the Heisenberg uncertainty principle.  This would indeed point towards a quantum nature of gravity.
\vskip 0.2 cm
\noindent\textit{\textbf{Summary: }}In this paper, we have considered a linearly polarized gravitational wave interacting with a gravitational wave detector. In the current model a quantized gravitational wave interacts with a detector where the detector variables obey the modified Heisenberg uncertainty relation (also known as the generalized uncertainty principle). In our analysis, we have considered that initially the gravitational wave was not interacting with the detector making it possible to write the initial state of the system as a tensor product state of the state corresponding to the graviton and the initial state corresponding to the detector. 
Following the approach in \cite{QGravNoise,QGravLett,QGravD}, we have summed over all the final states of the graviton. 
We then integrate out the coordinates and momenta corresponding to the graviton to obtain the influence functional involved in this entire process using a path integral approach. Following this one graviton analysis, we have extended our study to a gravitational field where the field modes are in coherent states and then squeezed states respectively. We finally obtain a stochastic Langevin-like equation including a noise term which can be considered as a quantum gravitational correction to the classical geodesic deviation equation. Our work focusses in the inclusion of a minimal length scale correction in the Heisenberg uncertainty principle, corresponding to the detector variables, leading to a GUP modified stochastic Langevin equation. We then obtain an approximate solution of the time dependent detector arm length  and calculate the corresponding standard deviation in it. Although in case of the field modes being in a coherent state leads to a minuscule correction to the classical geodesic deviation equation, for a squeezed state analysis we find that along with the noise term the GUP effect also gets an exponential boost due to the existence of a tunable squeezing parameter. It is surprising in a sense that the GUP effect is coming from the detector variables spanning the modified phase space only and therefore the squeezing embedded in the field states can amplify hidden signatures of the minimum length scale corrections considered in the detector. We then obtain a bound on the dimensionless GUP parameter which is tighter than the bounds obtained earlier using gravitational wave data. We finally plot the dimensionless standard deviation term due to the GUP effect with respect to time and observe that it may be possible to detect hidden GUP signatures while detecting primordial gravitational waves (with a squeezing parameter $r\simeq 44$) in future generation of gravitational wave detectors (LISA) along with the detection of gravitons. 

\section*{Acknowledgement}
\noindent We thank the referees for constructive and valuable comments which have helped us to improve the manuscript substantially.


\begin{thebibliography}{8}
\bibitem{QGravNoise}
M. Parikh, F. Wilczek, and G. Zahariade, \href{https://www.worldscientific.com/doi/abs/10.1142/S0218271820420018}{Int. J. Mod. Phys. D 29 (2020) 2042001}.
\bibitem{QGravLett}
M. Parikh, F. Wilczek, and G. Zahariade, \href{https://link.aps.org/doi/10.1103/PhysRevLett.127.081602}{Phys. Rev. Lett. 127 (2021) 081602}.
\bibitem{QGravD}
M. Parikh, F. Wilczek, and G. Zahariade, \href{https://link.aps.org/doi/10.1103/PhysRevD.104.046021}{Phys. Rev. D 104 (2021) 046021}.
\bibitem{String1}
D. Amati, M. Ciafaloni, and G. Veneziano, \href{https://doi.org/10.1016/0370-2693(89)91366-X}{Phys. Lett. B 216 (1989) 41}.
\bibitem{String2}
K. Konishi, G. Paffuti, and P. Provero, \href{https://doi.org/10.1016/0370-2693(90)91927-4}{Phys. Lett. B 234 (1990) 276}.
\bibitem{Loop1}
C. Rovelli,  \href{https://doi.org/10.12942/lrr-1998-1}{Living Rev. Relativ. 1 (1998) 1}.
\bibitem{Loop2}
S. Carlip, \href{https://iopscience.iop.org/article/10.1088/0034-4885/64/8/301}{Rep. Prog. Phys. 64 (2001) 885}.
\bibitem{Snyder}
H. S. Snyder, \href{https://link.aps.org/doi/10.1103/PhysRev.71.38}{Phys. Rev. 71 (1947) 38}.
\bibitem{Snyder2}
H. S. Snyder, \href{https://link.aps.org/doi/10.1103/PhysRev.72.68}{Phys. Rev. 72 (1947) 68}.
\bibitem{AC}
A. Connes. \textit{Noncommutative Geometry} (1994), Academic Press.
\bibitem{NCG1}
F. Girelli, E. R. Livine, and D. Oriti, \href{https://doi.org/10.1016/j.nuclphysb.2004.11.026.}{Nucl. Phys. B 708 (2005) 411}.
\bibitem{Guendelman}
E. I. Guendelman, \href{https://doi.org/10.1142/S0218271821420281}{Int. J. Mod. Phys. D 30 (2021) 2142028}.
\bibitem{BRON1}
M. P. Bronstein, Zh. Eksp. Teor. Fiz. 6 (1936) 195.
\bibitem{BRON2}
M. P. Bronstein, Phys. Z. Sowjetunion 9 (1936) 140.
\bibitem{MEAD}
C. A. Mead, \href{https://link.aps.org/doi/10.1103/PhysRev.135.B849}{Phys. Rev. B 135 (1964) 849}.
\bibitem{MAGGIORE}
M. Maggiore, \href{https://doi.org/10.1016/0370-2693(93)90785-G}{Phys. Lett. B 319 (1993) 83}.
\bibitem{SCARDIGLI}
F. Scardigli, \href{https://doi.org/10.1016/S0370-2693(99)00167-7}{Phys. Lett. B 452 (1999) 39}.
\bibitem{ADLERSAN}
R. J. Adler and D. I. Santiago, \href{https://www.worldscientific.com/doi/abs/10.1142/S0217732399001462}{Mod. Phys. Lett. A 14 (1999) 20}.
\bibitem{ADLERCHENSAN}
R. J. Adler, P. Chen, and D. I. Santiago, \href{https://doi.org/10.1023/A:1015281430411}{Gen. Relativ. Gravit. 33 (2001) 2101}.
\bibitem{RABIN}
R. Banerjee and S. Ghosh, \href{https://doi.org/10.1016/j.physletb.2010.04.008.}{Phys. Lett. B 688 (2010) 224}.
\bibitem{SG1}
S. Gangopadhyay, A. Dutta, and A. Saha, \href{https://doi.org/10.1007/s10714-013-1661-3}{Gen. Relativ. Gravit. 46  (2014) 1661}.
\bibitem{SCARDIGLI2}
F. Scardigli and R. Casadio, \href{https://doi.org/10.1140/epjc/s10052-015-3635-y}{Eur. Phys. J. C 75 (2015) 425}.
\bibitem{SG2}
R. Mandal, S. Bhattacharyya, and S. Gangopadhyay, \href{https://doi.org/10.1007/s10714-018-2468-z}{Gen. Relativ. Gravit. 50 (2018) 143}.
\bibitem{Ong}
Y. C. Ong, \href{http://dx.doi.org/10.1088/1475-7516/2018/09/015}{JCAP 09 (2018) 015}.
\bibitem{EPJC}
L. Buoninfante, G. G. Luciano, and L. Petruzzeillo, \href{https://doi.org/10.1140/epjc/s10052-019-7164-y }{Eur. Phys. J. C  79 (2019) 663}.
\bibitem{BMajumder}
B. Majumder, \href{https://doi.org/10.1016/j.physletb.2011.05.076}{Phys. Lett. B 701 (2011) 384}.
\bibitem{DAS1}
S. Das and E. C. Vagenas, \href{https://link.aps.org/doi/10.1103/PhysRevLett.101.221301}{Phys. Rev. Lett. 101 (2008) 221301}.
\bibitem{DAS2}
S. Das and E. C. Vagenas, \href{https://cdnsciencepub.com/doi/10.1139/P08-105}{Can. J. Phys. 87 (2009) 233}.
\bibitem{IVASP}
I. Pikovski, M. R. Vanner, M. Aspelmeyer, M. S. Kim, and \v{C}. Brukner, \href{https://doi.org/10.1038/nphys2262}{Nat. Phys. 8 (2012) 393}.
\bibitem{IVASP2}
P. Bosso, S. Das, I. Pikovski, and M. R. Vanner, \href{https://link.aps.org/doi/10.1103/PhysRevA.96.023849}{Phys.  Rev. A 96 (2017) 023849}.
\bibitem{KSP}
S. P. Kumar and M. B. Plenio, \href{https://link.aps.org/doi/10.1103/PhysRevA.97.063855}{Phys. Rev. A 97 (2018) 063855}.
\bibitem{OTM0}
S. Sen, S. Bhattacharyya, and S. Gangopadhyay, \href{https://iopscience.iop.org/article/10.1088/1361-6382/ac55ab}{Class. Quant. Grav.  39 (2022) 075020}.
\bibitem{SG3}
S. Gangopadhyay and S. Bhattacharyya, \href{https://link.aps.org/doi/10.1103/PhysRevD.99.104010}{Phys. Rev. D 99 (2019) 104010}.
\bibitem{SG4}
S. Bhattacharyya, S. Gangopadhyay, and A. Saha, \href{https://iopscience.iop.org/article/10.1088/1361-6382/abac45}{Class. Quant. Grav. 37 (2020) 195006}.
\bibitem{OTM}
S. Sen, S. Bhattacharyya, and S. Gangopadhyay, \href{https://doi.org/10.3390/universe8090450}{Universe 8 (2022) 450}.
\bibitem{Petruzzeillo}
L. Petruzzeillo and F. Illuminati, \href{https://doi.org/10.1038/s41467-021-24711-7}{Nat. Commun. 12 (2021) 4449}.
\bibitem{DasModak}
S. Das and S. K. Modak, \href{http://dx.doi.org/10.1088/1361-6382/ac38d3}{Class. Quant. Grav. 39 (2022) 015005}.
\bibitem{Farag}
A. F. Ali, S. Das, E. C. Vagenas, \href{https://doi.org/10.1016/j.physletb.2009.06.061}{Phys. Lett. B 678 (2009) 497}.
\bibitem{Farag2}
A. F. Ali, S. Das, E. C. Vagenas, \href{https://link.aps.org/doi/10.1103/PhysRevD.84.044013}{Phys. Rev. D 84 (2011) 044013}.
\bibitem{SGSB}
S. Gangopadhyay and S. Bhattacharyya, \href{https://link.aps.org/doi/10.1103/PhysRevD.104.026003}{Phys. Rev. D 104 (2021) 026003}.
\bibitem{Kempf}
A. Kempf, G. Mangano, and R. Mann, \href{https://link.aps.org/doi/10.1103/PhysRevD.52.1108}{Phys. Rev. D 52 (1995) 1108}.
\bibitem{FeynmanVernon}
R. P. Feynman and F. L. Vernon, Jr., \href{https://www.sciencedirect.com/science/article/pii/000349166390068X}{Ann. Phys. 24 (1963) 118}.
\bibitem{Douglas}
M. Bishop, J. Contreras, and D. Singleton, \href{https://doi.org/10.1142/S0218271822410024}{Int. J. Mod. Phys. D 31 (2022) 2241002}.
\bibitem{KSThorne}
K. S. Thorne, \href{https://ui.adsabs.harvard.edu/abs/1969ApJ...158..997T}{Astrophys. J. 158 (1969) 997}.
\bibitem{WLBurke}
W. L. Burke, \href{https://doi.org/10.1063/1.1665603}{J. Math. Phys. (N.Y.) 12 (1971) 401}.
\bibitem{Chandrasekhar}
S. Chandrasekhar and F. Paul Esposito, \href{https://ui.adsabs.harvard.edu/abs/1970ApJ...160..153C}{Astrophys. J. 160 (1970) 153}.
\bibitem{Sasaki}
Y. Mino, M. Sasaki, and T. Tanaki, \href{https://link.aps.org/doi/10.1103/PhysRevD.55.3457}{Phys. Rev. D 55 (1997) 3457}.
\bibitem{Gradshteyn}
I. S. Gradshteyn and I. M. Ryzhik, \textit{Table of Integrals, Series and Products} (Academic Press, Amsterdam, 2007), 6th ed.
\bibitem{Accadia}
T. Accadia \textit{et. al.}, \href{https://dx.doi.org/10.1088/1748-0221/7/03/P03012}{J. Inst. 7 (2012) P03012}.
\bibitem{SukantaDaSunandan}
S. Bhattacharyya, S. Gangopadhyay, and A. Saha, \href{https://dx.doi.org/10.1088/1361-6382/abac45}{Class. Quant. Grav. 37 (2020) 195006}.
\bibitem{GWaveDas}
A. Das, S. Das, N. R. Mansour, and E. C. Vagenas, \href{https://doi.org/10.1016/j.physletb.2021.136429.}{Phys. Lett. B 819 (2021) 136429}.
\bibitem{PI1}
A. Albrecht, P. Ferreira, and M. Joyce, and T. Prokopec, \href{https://journals.aps.org/prd/abstract/10.1103/PhysRevD.50.4807}{Phys. Rev. D 50 (1994) 4807}.
\bibitem{PI2}
L. P. Grishchuk and Y. V. Sidorov, \href{https://link.aps.org/doi/10.1103/PhysRevD.42.3413}{Phys. Rev. D 42 (1990) 3413}.
\bibitem{PI3}
B. L. Hu, G. Kang, and A. Matacz, \href{https://doi.org/10.1142/S0217751X94000455}{Int. J. Mod. Phys. A 09 (1994) 991}.
\bibitem{Soda}
S. Kanno, J. Soda, and J. Tokuda, \href{https://link.aps.org/doi/10.1103/PhysRevD.103.044017}{Phys. Rev. D 103 (2021) 044017}.
\end{thebibliography}
\end{document}